\documentclass[twoside,12pt]{article} 

\usepackage{epsf}

\def\RecPlay{{\sc RecPlay}}
\def\Jiti{JiTI}
\def\topline{\vspace*{-1mm}\rule{\textwidth}{0.1pt}\vspace*{5mm}}
\def\bottomline{\vspace*{0mm}\rule{\textwidth}{0.1pt}\vspace*{-4mm}} 
\newcommand{\cc}[1]{\multicolumn{1}{c}{#1}}

\begin{document} 
\pagestyle{myheadings} 
\markboth{AADEBUG 2000}{Non-intrusive on-the-fly data race detection using execution replay.} 
\title{Non-intrusive on-the-fly data race detection using execution replay.\footnote{In M. Ducass\'e (ed), proceedings of the
Fourth International Workshop on Automated Debugging (AADEBUG 2000),
August 2000, Munich. COmputer Research Repository
(http://www.acm.org/corr/), cs.SE/0011005; whole proceedings:
cs.SE/0010035.}}

\author{Michiel Ronsse and Koen De Bosschere\\
Department of Electronics and Information Systems\\
Ghent University, Belgium\\
ronsse@elis.rug.ac.be}

\date{} 

\maketitle 

\begin{abstract}
This paper presents a practical solution for detecting data races in
parallel programs. The solution consists of a combination of execution
replay (\RecPlay) with automatic on-the-fly data race detection. This
combination enables us to perform the data race detection on an
unaltered execution (almost no probe effect).  Furthermore, the usage
of multilevel bitmaps and snooped matrix clocks limits the amount of
memory used. As the record phase of \RecPlay\ is highly efficient,
there is no need to switch it off, hereby eliminating the possibility
of Heisenbugs because tracing can be left on all the time.
\end{abstract}

\section{Introduction} %%%%%%%%%%%%%%%%%%%%%%%%%%%%%%%%%%%%%%%%%%%%%%%%%%

The never ending urge for faster and more robust computers, combined
with the existence of cheap processors causes a proliferation of
inexpensive multiprocessor machines. Multithreaded applications are
needed to exploit the full processing power of these machines, causing
a widespread use of parallel applications. Even most contemporary
applications that are not CPU-intensive are multithreaded because the
multithreaded paradigm makes it easier to develop servers,
applications with an MDI ({\em multiple document interface}) such as
Windows programs, etc.

However, developing multithreaded programs for these machines is not
easy as it is harder to get a good view on the state of a parallel
program. This is caused by the fact that there are a number of
threads\footnote{In this paper we will consider an execution of a
parallel program as being a process consisting of $N$ threads
executing on a machine with $N$ processors.}  running
simultaneously. Moreover, the very fact that the computation is split
into simultaneous parts can introduce errors that do not exist in
sequential programs. These {\em synchronisation errors} show up
because parallel programs are developed in order to let a number of
threads work on the same problem, hence they will share data. It is of
paramount importance that the accesses performed by the different
threads are properly synchronised.  Too much synchronisation will
deadlock the program, while a lack of synchronisation will lead to
race conditions. Such a {\em race condition} occurs when two or more
concurrently executing threads access the same shared memory location
in an unsynchronised way, and at least one of the threads modifies the
contents of the location.

\begin{figure}
\topline
\begin{verbatim}
#include <pthread.h>

unsigned global=5;

thread1(){
  global=global+6;
}

thread2(){
  global=global+7;
}

main(){
  pthread_t t1, t2;

  pthread_create(&t1, NULL, thread1, NULL);
  pthread_create(&t2, NULL, thread2, NULL);

  pthread_join(t1, NULL);
  pthread_join(t2, NULL);
}                          
\end{verbatim}
\bottomline
\caption{A program containing a race condition on {\tt global}.}
\label{sol-vb}
\end{figure}    

The program depicted in Figure~\ref{sol-vb} is an example of a program
containing a race condition. The program uses a shared global variable
({\tt global}), initialised with value 5. The first thread ({\tt
main}) starts two additional threads, {\tt t1} and {\tt t2}, executing
respectively the functions {\tt thread1} and {\tt thread2}. Both
threads change the global variable. As the threads do not synchronise,
the final value of {\tt global} will be 18, 11 or 12. The latter two
are the result of one of the threads overwriting the new value written
by the other thread.

As race conditions are (most of the time) considered bugs, they should
be removed. Unfortunately, race conditions are difficult to find because
their occurrence depends on small timing variations. Although it is
possible to detect race conditions using a static approach (using source
code analysis, as done by \cite{locklint}) this is not feasible for
nontrivial programs (it has been shown that detecting race conditions
statically is an NP complete problem for programs that use
synchronisation that is powerful enough to implement mutual exclusion
\cite{one_sema, Netzer_DR}). Therefore, most race detection tools
detect race conditions dynamically during an actual program execution.

In this paper, we present and describe \RecPlay \cite{mr-tocs}, a tool
that detects data races (data races are harmful race conditions; see
next section) dynamically. \RecPlay\ was implemented for the Solaris
operating system running on Sun multiprocessors with SPARC
processors. During the test phase of \RecPlay, we have found data
races in several programs we tested (including the SPLASH-2
suite~\cite{splash2}, and the Athapascan
system~\cite{athapascan}). All of the races that were reported were
genuine data races that had stayed undetected until then.

In the next section, we start with a few definitions about race
conditions followed by a description of data race detection
techniques. The section ends with an overview of our method. The next
section describes our implementation. This part is followed by
an evaluation section containing performance data. The paper is
concluded with an overview of related work.

%%%%%%%%%%%%%%%%%%%%%%%%%%%%%%%%%%%%%%%%%%%%%%%%%%%%%%%%%%%%%%%%%%%%%%%%%%
\section{Data races}\label{defs}

\subsection{Definitions}

A {\em race condition} occurs when two threads access the same shared
location in an unsynchronised way and at least one access modifies the
value at the location. We distinguish two types of race conditions:
race conditions that are used to make a program intentionally
nondeterministic: {\em synchronisation races}, and race conditions
that were not intended by the programmer ({\em data races}).

We need {\em synchronisation races} to allow for competition between
threads to enter a critical section, to lock a semaphore, or to
implement load balancing.  Removing synchronisation races makes a
program completely deterministic. Therefore, in this paper, we do not
consider synchronisation races a programming error, but a functional
and useful characteristic of a parallel program.

{\em Data races} are not intended by the programmer, and are most of
the time the result of improper synchronisation. By changing the
synchronisation, data races can (and should) always be
removed\footnote{There are a few applications that use data races
intentionally, e.g.\ relaxation algorithms.}.

It is important to notice that the distinction between a data race and
a synchronisation race is actually a pure matter of abstraction. At
the implementation level of the synchronisation operations, a
synchronisation race is caused by a genuine data race (e.g., spin
locks, polling, etc.) on a synchronisation variable.

If the programmer creates his/her own synchronisation operations, we
also have to make a distinction between {\em feasible} races and
{\em apparent} races. The difference between them can be
intuitively understood from the following example.
\begin{center}\tt
\begin{tabular}{ll}
\hline
thread$_1$ & thread$_2$ \\ 
\hline
result=x; & while (done==FALSE); \\
done=TRUE; & y=result; \\
\hline
\end{tabular}
\end{center}
Technically speaking, in this fragment, there are no explicit
synchronisation operations, and therefore, a naive race detection will
detect two races: one on {\tt result} and one on {\tt done}. However,
after taking the semantics of the program fragment into account, there
is only one data race that can actually occur, namely on {\tt done}
because the data race on {\tt result} cannot occur, provided that {\tt
done} was set to {\tt FALSE} before the two threads were
created.\footnote{This is true if the computer has a memory model that
guarantees at least processor consistency. Modern processors with
weaker memory models require some kind of serialisation instruction
(e.g.\ a {\tt STBAR} instruction on a SPARC) between the two store
operations performed by the first thread to make this work
correctly. This example only serves to explain the concept.}  We
say that his program fragment contains two {\em apparent} data races,
but only one data race is {\em feasible}.\footnote{All synchronisation
races should be feasible races. A synchronisation race that is
apparent but not feasible is an indication that the synchronisation is
superfluous.}

\subsection{Detecting data races dynamically} %%%%%%%%%%%%%%%%%%%%%%%%%%%%%%%

As mentioned before, a dynamic data race detector finds data races
that occur during a particular execution. In order to detect the
conflicting memory operations, one should collect all memory
operations executed during a particular execution and information
about their concurrency. The concurrency depends on the (order of the)
executed synchronisation operations. By analysing the traced
information one can find the data races that occurred during that
particular execution. However, this type of post-mortem approach is
not feasible due to the very large trace files that have to be
generated. However, by carefully examining the requirements for data
race detection it is possible to limit the information that has to be
traced without sacrificing the efficiency of the detection.

There exists basically two (dual) methods for detecting data races using
collected memory and concurrency information:
\begin{itemize}
\item for each access to a global variable, the access is compared against
previous accesses by other threads to the same variable. This requires
us to collect, for each global variable, information about past
accesses\footnote{It suffices to collect information about the last
load and store operation of each thread.}. It is obvious that this will
lead to a huge memory consumption, especially if each memory location
is a potential global variable with life time equal to that of the
program itself, as is the case for e.g.\ programs written in the $C$ language.
\item for all parallel pieces of code, the memory operations are collected and
compared. This method consolidates the fact that all memory operations between
two successive synchronisation operations (called segments from now on) satisfy
the same concurrency relation: they are either all concurrent or not concurrent
with a given operation.

Given the sets $L(i)$ and $S(i)$ containing the addresses used by the load and store
operations in segment $i$, the concurrent segments $i$ and $j$ contain racing
operations if and only if
\[
\Big(\big(L(i) \cup S(i)\big) \cap S(j)\Big)  \cup \Big(\big(L(j) \cup S(j)\big) \cap S(i)\Big)  \neq \emptyset
\]
Therefore, data race detection basically boils down to collecting the
sets $L(i)$ and $S(i)$ for all segments executed and comparing
parallel segments.
\end{itemize}

It is clear that the second method is better suited for programming languages
with unconstrained life time of and access to shared variables. However, two
problems remain:

\begin{enumerate}
\item treating all memory accesses the same way prevents us from making a
distinction between the synchronisation (due to the memory accesses
performed by the synchronisation operations) and the data races;
\item collecting all memory operation introduces a huge overhead, both in time
and space. The unavoidable intrusion gives rise to the probe effect\cite{Gait2}, possibly
causing Heisenbugs.
\end{enumerate}

The last problem is a dramatic problem, as it seems unsolvable. Indeed, we
cannot circumvent the tracing of all memory operations if one wants to detect
data races. However, it is possible to alleviate the problem by tracing an
application in two phases.

\subsection{Data race detection using two phases}
Data race detection using two phases detects data race in a programn
execution using two `equivalent' executions:
\begin{enumerate}
\item synchronisation races are dealt with during the first
phase. During this {\em record} phase a limited amount of information
about the execution is traced;
\item during the second phase, all the information needed to detect data races
is traced, and the information traced during the first phase is used in order to
guide the second execution. This execution is guided in order to enforce the
execution to be {\em equivalent} (two executions are said to be {\em equivalent}
if they yield the same internal program flow) with the first execution.
\end{enumerate}

In order to be able to force an execution to be equivalent with another
execution, information about the `decisions' that (could) have an
impact on the program flow of the first execution should
be traced. This information is:
\begin{itemize}
\item input received from sources outside the process such as reading 
from disk, system calls such as {\tt gettimeofday()}, {\tt
random()}, \ldots This paper will assume that all input can be fed back during
a replayed execution, either because the input is on stable disk or is
replayed, e.g. by intercepting system calls;
\item `input' received from other threads that are part of the same
process\footnote{In this paper, we only deal with parallel threads
belonging to one process. Different processes exchanging data (e.g.\
using {\tt mmap()}) are not considered.}. Indeed, as each load
operation possibly reads a value written by another thread this
should be considered input. Note that these are the race conditions
mentioned before.
\end{itemize}

It seems that this does not solve our problem: we have to trace all
memory operations in order to get a faithful re-execution, making it
impossible to get a trace of an execution that is not
perturbed. However, tracing {\em all} race conditions is not
necessary. If we only trace the synchronisation races, a correct
re-execution is only possible in the absence of data races. If the
execution contains one or more data races, the re-execution will
(possibly) fail after the data race starts having an effect on the
execution. As the point at which the data race starts effecting the
execution occurs {\em after} the occurrence of the data race itself,
it is possible to detect the data race during the guided re-execution.

Using two phases has a number of benefits:
\begin{itemize}
\item the overhead introduced during the record phase will be small as:
\begin{itemize}
\item the number of synchronisation operations is usually much smaller than
the total number of memory operations,
\item we can trace a synchronisation operation as being a single operation, although
the synchronisation operation will typically perform (e.g.\ in a
spinning loop) a number of memory operations,
\item making a distinction from the beginning between synchronisation races
and data races makes it easier to detect only the data races
during the replay phase,
\end{itemize}
\item it is easy to trace the synchronisation operations as they are
usually implemented in a dynamically linked library, making it
possible to intercept them by interposing an instrumented library.
\end{itemize}

This solution allows us to perform an intrusive data race detection
during a `normal' re-execution. The price we have to pay is that we
need two executions in order to collect all the information.

Although this solves part of our problem (the intrusion is limited),
this is not a viable solution as we still need to collect a lot of
information during the replayed execution. However, it is possible to
limit the amount of information we need to collect in other to find
data races.

\subsection{Limiting the amount of traced information}

A naive tool could perform the data race detection post-mortem,
forcing us to collect all information in memory or on disk. However,
if an execution contains a data race, there is no need to keep looking
for other data races. Indeed, due to the so-called avalanche effect,
the first data race should be removed before the others. Moreover, as
soon as (the effects of) a data race occurs, the replayed execution is
no longer guaranteed to be correct. On-the-fly detection is therefore
a viable alternative: data race detection is performed during the
execution, and as soon as a data race is found, information about it
is displayed. This method will use less memory if a data race shows up
near the start of the program. However, a program execution with a
data race near the end, or with no data race at all, will still use a
huge amount of memory as this approach keeps on adding $L(i)$ and
$S(i)$ sets to memory. It is therefore important to remove them as
soon as it is clear that there are no more segments that will need
them for future comparison. A segment can be removed if all other
threads have progressed beyond the point in logical time when the
segment did terminate. After that point in time, there can be no more
segments started that possibly cause a data race.

If a program execution contains a data race, the programmer is
interested in the racing instructions and the variable they race
on. Therefore, we have to collect the address of the instruction and
the address of the variable used for every memory operation contained
in a segment. However, as this amounts to a huge memory consumption,
\RecPlay\ only stores the addresses used and no information about the
instructions themselves. This information is sufficient to detect data
races, but not the offending instructions. A third execution ({\em
identification phase}) is needed to locate these instructions. During
this execution, the knowledge about the data race (address of the
variable, identities of the threads, the identity of the segment and
type of operation (load or store)) is sufficient to locate the actual
racing instructions.

The online data race detection is therefore performed by \RecPlay\
using three phases:
\begin{enumerate}
\item record phase: trace all synchronisation operations,
\item replay \& detect phase:
\begin{itemize}
\item for each executed memory instruction: trace the address used,
\item for each synchronisation operation executed:
\begin{itemize}
\item make sure it is executed at the right moment: between the same
two synchronisation operations as during the recorded execution,
\item compare the traced memory operations of the ended segment with previously
executed parallel segments. If a data race is found: end the program,
\item add the collected information ($L(i)$ and $S(i)$) to the list of segments,
\item remove obsolete segments from the list,
\end{itemize}
\end{itemize}
\item if a data race was found: identification phase.
\end{enumerate}                            

In the next section, the techniques used to implement \RecPlay\ are
discussed.

%%%%%%%%%%%%%%%%%%%%%%%%%%%%%%%%%%%%%%%%%%%%%%%%%%%%%%%%%%%%%%%%%%%%%%%%%%%%%%
\section{Implementation}

As mentioned in the introduction, our method has been implemented for
Solaris. Our tool \RecPlay\ will detect the first feasible data race
in an execution as long as the application under test only uses the
Solaris (or POSIX) synchronisation operations. This still applies if
one builds own synchronisation operations (or uses a third party
library) on top of the Solaris or POSIX variants. 

Data races detected in user programs that also use their own
synchronization operations can also be {\em apparent} data races that
are not feasible.  However, at least one of the apparent data races
must be a feasible data race~\cite{Netzer_DR} that is used to
synchronize the other memory references (and hence is a
synchronization race). By replacing user defined synchronization
operations by Solaris synchronization operations, or by putting the
new synchronization operations in a separate library that is flagged
as synchronization library, we can prevent \RecPlay\ from detecting
such synchronization races during subsequent runs.

\subsection{Instrumentation}

In order to collect the memory reference information and to
record/replay the synchronisation operations, \RecPlay\ intercepts
these operations using \Jiti\ (Just in Time Instrumentation,
\cite{MR_JITI}). \Jiti\ was developed for \RecPlay\ but is a general
instrumentation technique able to deal with hard to instrument
features such as code in data, data in code and self modifying code.

%%%%%%%%%%%%%%%%%%%%%%

The two major difficulties when inserting code into binaries are (i) correctly
distinguishing between code and data (especially when code is located in data,
when data is located in code, or when self-modifying code is used) and (ii)
correctly relocating the code and data after inserting instrumentation code.

In existing systems, these two difficulties can only be solved by
applying a sophisticated analysis (disassembling the program into {\em
basic blocks} and {\em control flow graphs}) of the binary. Hereby,
assumptions have to be made about the origin of the code. Most systems
can be broken by offering hand-written machine code to it. A careful
analysis for large programs can take a inordinate amount of resources
(both in time and space).

\Jiti\ solves these problems (i) by creating two versions of
the process: one for the data accesses and one for the code accesses
and (ii) by not inserting instrumentation code in the process, but by
replacing instructions by calls to instrumentation code.  By cloning a
process and by executing the code from an (instrumented) copy, and
using the data from the other copy, we get rid of the need to
distinguish between code and data.

Given the size $\alpha$ of the original process, a clone is created at
address $\delta$ ($>\alpha$) (up to address $\delta+\alpha$). Since we
do not insert instructions, but only replace instructions, the
instrumented version of the instruction at address $i$ will reside at
address $\delta+i$.

We now have access to two copies: one copy that is entirely considered
as data, and another copy that is entirely considered as code. In
order to make sure that data is taken from one copy, and the code from
the other copy, we will have to modify addresses that do not point to
the right copy (by increasing or decreasing them with $\delta$). It
turns out that addresses used to access data are mostly absolute data
addresses, while, on contemporary microprocessors, code addresses are
mostly relative addresses (position independent code). Since relative
addresses do not require relocation when moved to another location in
memory, the best choice is to fetch the code from the clone, and the
data from the original copy of the process.  This means that the
relocation effort can be limited to the rare --especially on a SPARC
processor-- absolute code addresses (to make sure that the execution
will never jump back to the non-instrumented version of the
process\footnote{Some instructions that are known to make use of
correct absolute addresses (such as returns), do not need to be
instrumented.})  and to the rare code-relative data addresses that
might be used in the code (e.g., for data that is located in the code,
such as address tables).

In these cases where the relocation cannot be done {\em at clone time}
(e.g.\ for memory operations, where it is impossible to distinguish
the memory operations that use relative addresses from those that use
absolute addresses), it suffices to check the addresses {\em
on-the-fly} by instrumenting the instructions that make use of them.
The instrumentation of the clone can be performed before the program
starts or during the execution (by filling the clone with trapping
instructions and by instrumenting an instruction (or a whole page)
whenever an instruction is executed for the first time). It is even
possible to have different clones (for parallel programs) with
different instrumentation for different threads.  It is also possible
to change the instrumentation during the execution. This can be used
to deal with self-modifying code: store instructions are instrumented
to write the data to the original and a trap instruction to the
clone. If the trap instruction is ever executed, \Jiti\ intercepts the
trap and instruments the instruction (if necessary). By having the
instrumentation code do further instrumentation, the instrumentation
can modify itself dynamically.

%%%%%%%%

\Jiti\ is also able to intercept
dynamically linked library calls, and does not instrument the memory
operations in these libraries. This is ideal for making a distinction
between synchronisation and data races: all Solaris synchronisation
operation are contained in a dynamically linkable
library.\footnote{The library has no static variant, as is the case
for most of the system libraries.}  As such, the many memory
operations that are used by the synchronisation operations are not
seen.

\RecPlay\ and \Jiti\ have been implemented for Sun multiprocessors
running Solaris 7. The implementation uses the dynamic linking and
loading facilities of the operating system. A number of dynamic
loadable libraries have been created: one records the order of the
synchronisation operations, another one detects data races during a
replayed execution, and the last library identifies the offending
references while replaying a previously recorded run. Using an
environment variable, the Solaris program loader is forced to load one
of these libraries each time it loads a user program. The libraries
contain the necessary routines to perform initialisation before the
actual program starts. Neither the user program nor the thread library
needs to be modified as the \RecPlay\ library inserts itself
automatically between the user program and the thread library. This is
because we want to keep the \RecPlay\ operation (record, replay, race
detection, reference identification) a feature of an execution, rather
than a feature of a program.

%%%%%%%%%%%%%%%%%%%%%%%%%%%%%%%%%%%%%%%%%%%%%%%%%%%%%%%%%%%%%%%%%%%%%%%%%%
\subsection{Tracing and replaying the order of the synchronisation operations} 

In previous work, we have developed ROLT (Reconstruction of Lamport
Time\-stamps), an ordering-based record/replay method, where only the
partial order of synchronisation operations is traced by means of
(scalar) Lamport clocks attached to these operations. To get a
faithful replay, it is sufficient to stall each synchronisation
operation until all synchronisation operations with a smaller
timestamp have been executed. The execution traces consist of a
sequence of timestamps, one trace per thread. The method
\cite{LL94-01,MR-08} has the advantage that it produces small trace files
(it is not necessary to log all timestamps) and that it is less
intrusive than other existing methods~\cite{Netzer93}. Moreover, the
method allows for the use of a simple compression
scheme~\cite{MR95-03} which can further reduce the trace files.

%%%%%%%%%%%%%%%%%%%%%%%%%%%%%%%%%%%%%%%%%%%%%%%%%%%%%%%%%%%%%%%%%%%%%%%%%%%%%%%
\subsection{Collecting and comparing the addresses of the memory operations}

As mentioned before, two lists of addresses are used for every segment
$i$: the load operations are collected in $L(i)$ and the store
operations in $S(i)$. 

\begin{figure}
\begin{center}
\leavevmode
\epsfxsize=10cm\epsfbox{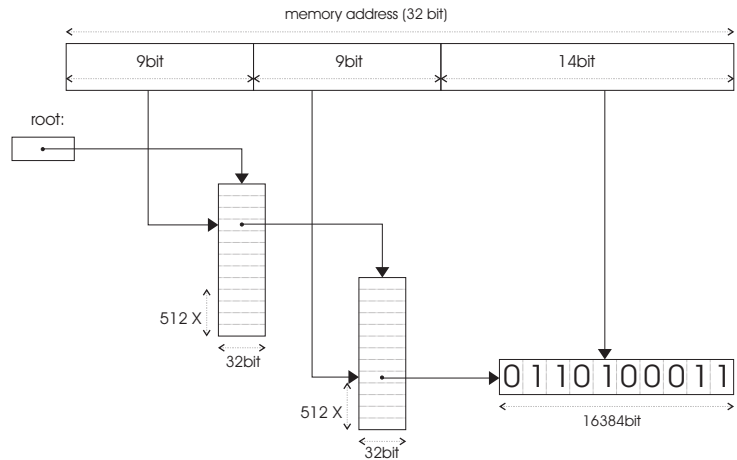}
\end{center}
\caption{A multilevel bitmap allows us to collect access information using a moderate amount of memory.}
\label{label:bitmap}
\end{figure}

These addresses are collected in a bitmap (one for each list). Such a
bitmap contains a $1$ on place $i$ if address $i$ was used. As a
linear bitmap would require $2^{32}/8=512$ MB, a multilevel bitmap is
used (see Figure~\ref{label:bitmap}).

The multilevel bitmap uses three levels: two levels containing page
tables and one level containing the actual bitmaps. The highest level
is addressed using the 9 highest bits of the addresses. This level
contains pointers to page tables at the second level. These page
tables are indexed using the next 9 bits and contain pointers to the
actual bitmaps. The actual bitmaps are indexed using the remaining 14
bits of the address. Indexing the multilevel bitmaps using a 9/9/14
scheme results in page tables and bitmaps of 2kB. This makes the
frequent allocating/deallocating of these data structures simple
enough to implement it ourselves without reverting to {\tt
malloc()/free()}. Measurements show that this speeds up an execution
with about $10\%$.

Although the use of a bitmap means that some information is lost --
two or more accesses of the same address are counted as one -- this is
not a problem for data race detection. Indeed, if one of these
accesses is involved in a race, the subsequent accesses are also
involved in a data race. The use of multilevel bitmaps makes it
possible to compare the segments in an efficient way.

%%%%%%%%%%%%%%%%%%%%%%%%%%%%%%%%%%%%%%%%%%%%%%%%%%%%%%%%%%%%%%%%%%%%%
\subsection{Detecting and comparing parallel segments}
\begin{figure}
\begin{center}
\leavevmode
\epsfysize=4.5cm\epsfbox{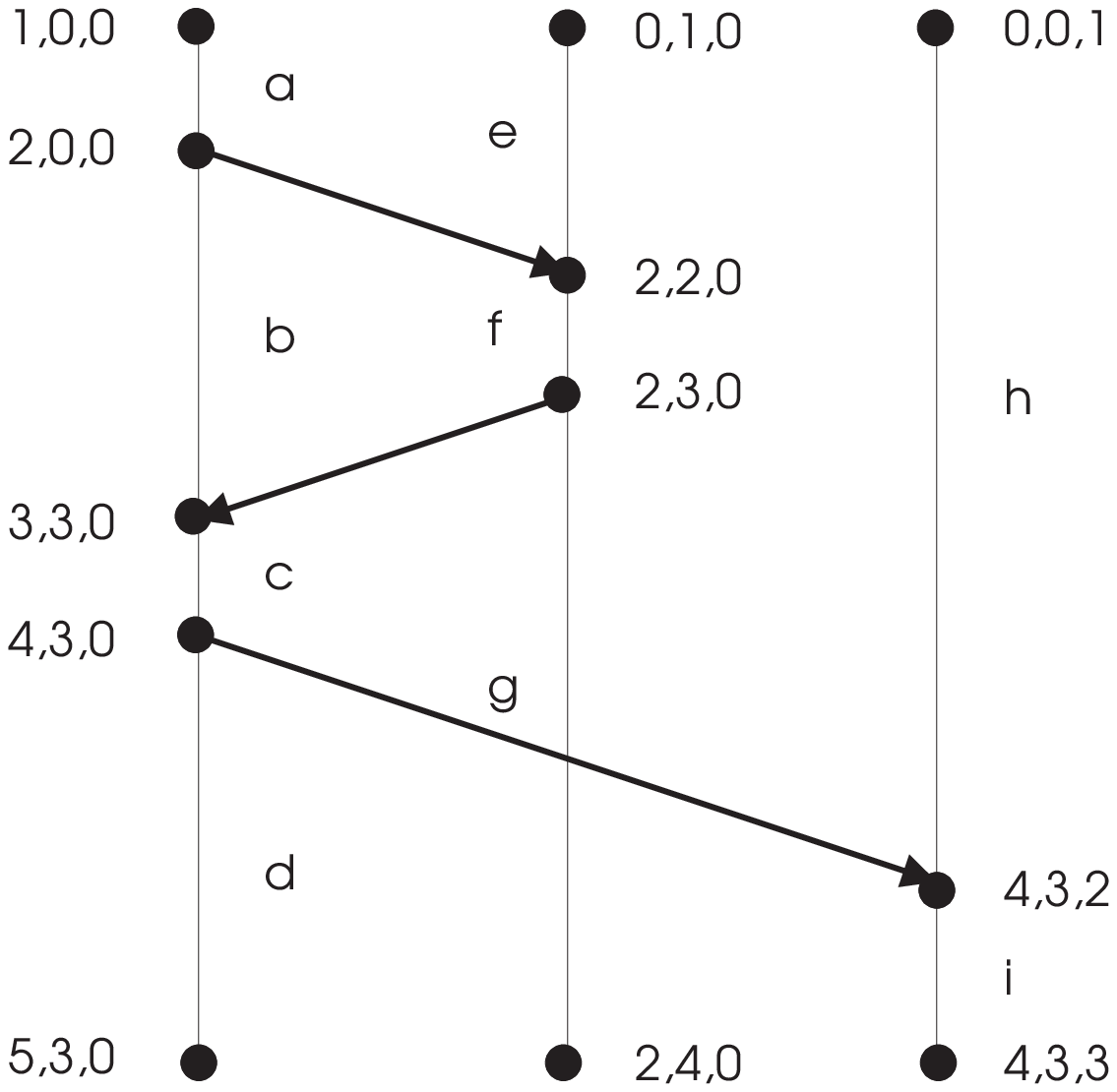}\qquad\qquad
\epsfysize=4.5cm\epsfbox{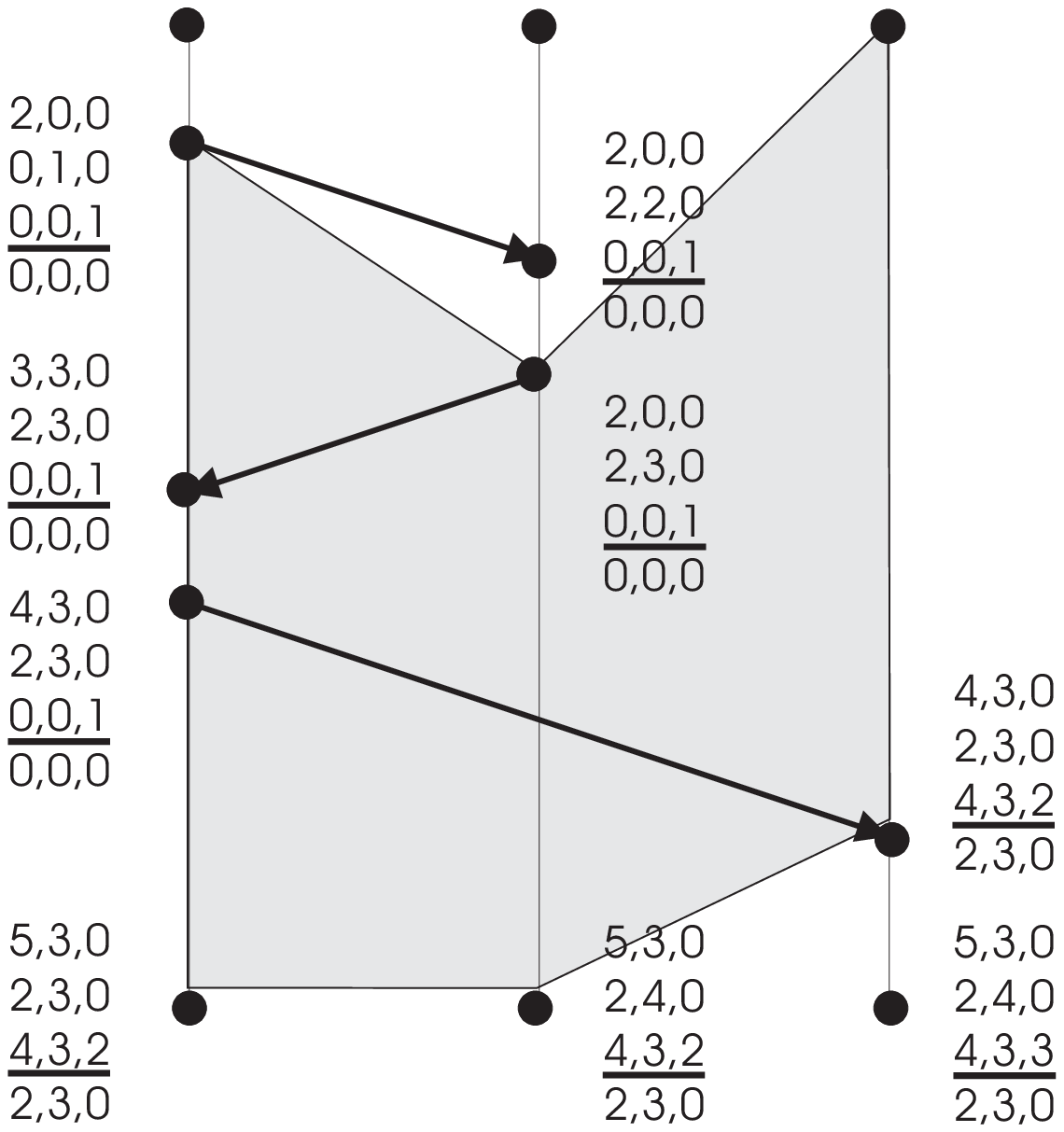}
\end{center}
\caption{Updating vector clocks and detecting data races:
vector clocks to detect concurrent segments (left part), logical
matrix clocks to discard $L$ and $S$ sets (right part). The right
figure shows a shaded area. The lower line of this area is a
representation of the progress of the execution (the points up to
where the threads have progressed). The upper line is the visible past
for the threads. All segments above that line can safely be
discarded. The shaded area shows the `current' segments.}
\label{vc}
\end{figure}

In our race detection tool, we use a classical logical vector
clock~\cite{Mattern,fidge} to detect concurrent segments
(left part of Figure~\ref{vc}). Updating vector clocks is time-consuming, but
this is not an issue during replay. As vector clocks are able to
represent the {\em happened-before} \cite{LamClock} relation (they are
strongly consistent), two vector clocks that are not ordered must
belong to concurrent segments. This gives us an easy way to detect
concurrent segments.

%%%%%%%%%%%%%%%%%%%%%%%%%%%%%%%%%%%%%%%%%%%%%%%%%%%%%%%%%%%%%%%%%%%%%%%%%%
\subsection{Detecting Obsolete Segments}

Logical matrix clocks~\cite{raynal96} are traditionally used to
discard information in a distributed environment: the componentwise
minimum of the columns yields the maximum number of segments per
thread that can be discarded (right part of
Figure~\ref{vc})~\cite{wuu,sarin}.

However, in practice we can discard more information than is indicated
by logical matrix clocks as logical clocks capture causality, which is
one of the weakest forms of event ordering. In a particular execution,
all events are executed is some order (not specified by the program),
even if they do not have a causal relationship. Classical logical
clocks are not able to capture this kind of additional
execution-specific ordering.

Our data race detection method uses {\em clock snooping}
\cite{mr-ispan}. Instead of maintaining matrix clocks, a {\em snooped matrix clock }is
built each time a segment ends, using the latest vector clock of the
processors. All segments that have a vector clock smaller than the
componentwise minimum of the snooped matrix clock can be discarded as
they will not be needed in the future anymore.

\begin{figure}[htbp]
\begin{center}
\leavevmode
\epsfysize=4cm\epsfbox{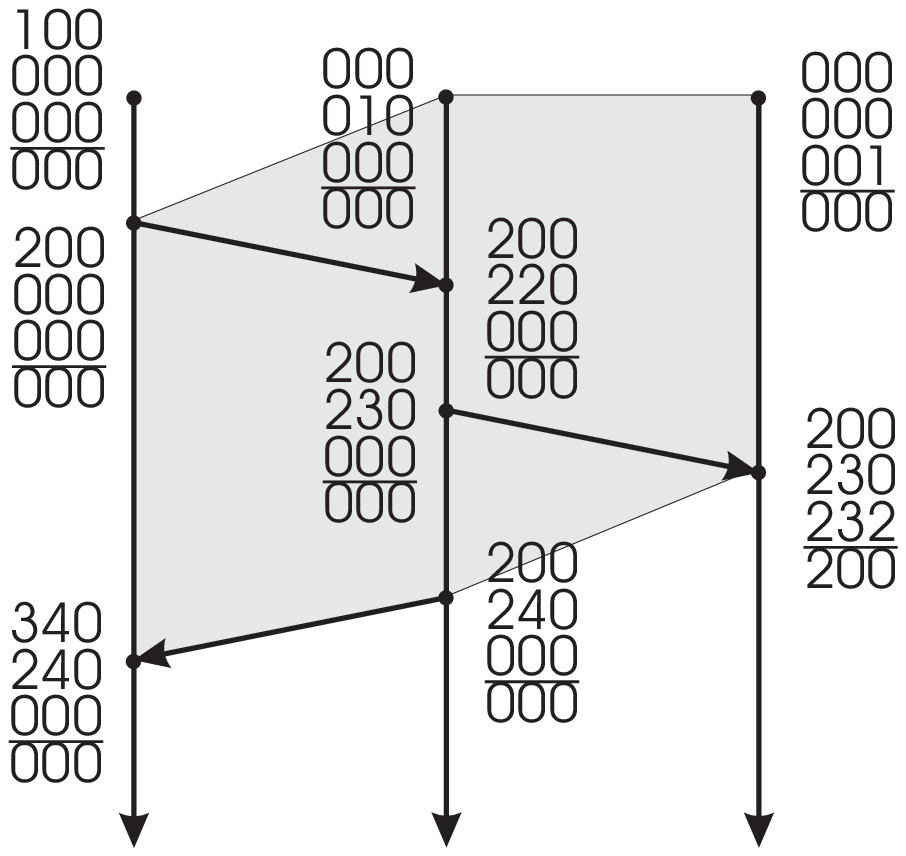}\qquad\qquad
\epsfysize=4cm\epsfbox{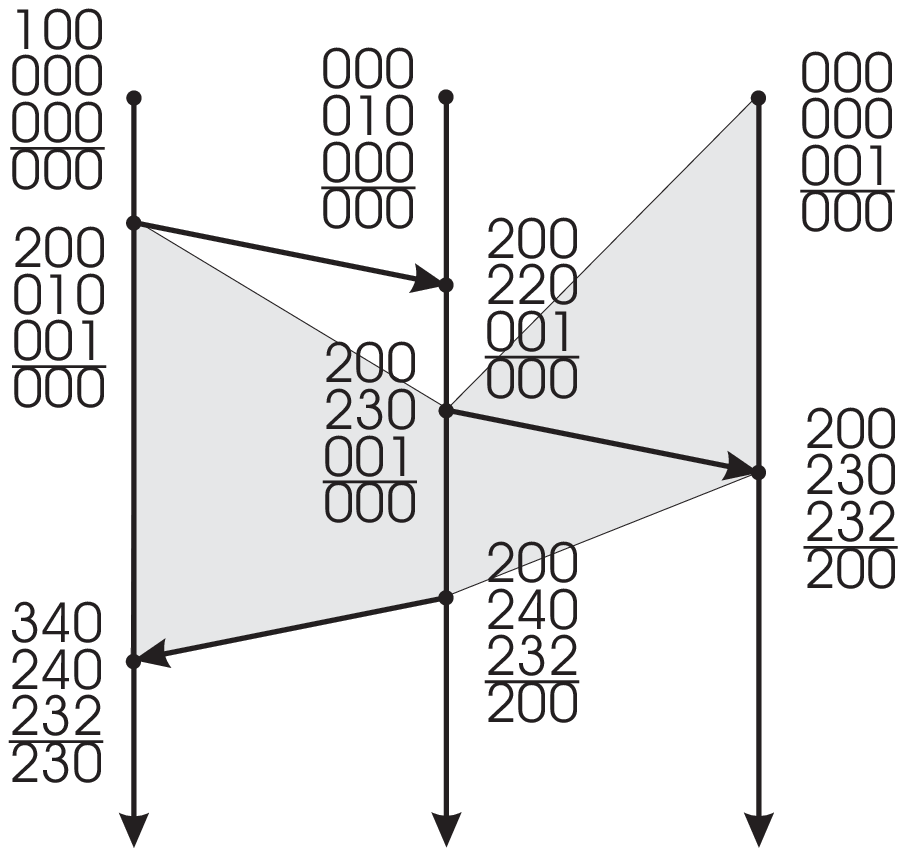}
\end{center}
\caption{Discarding information with logical (left) and snooped
(right) matrix clocks. The shaded consisting of the `current' segment
is clearly smaller when using snooped matrix clocks.}
\label{racedetection}
\end{figure}

\begin{figure}[htbp]
\begin{center}
\leavevmode
\epsfysize=7cm\epsfbox{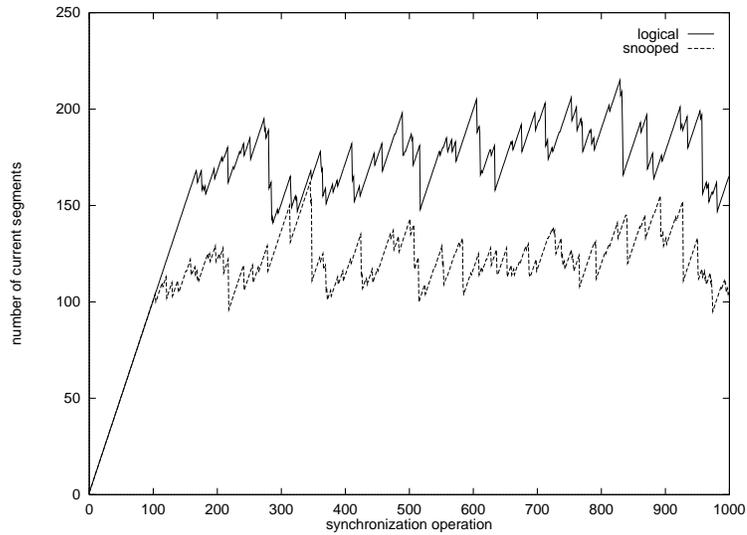}
\end{center}
\caption{Number of `current' segments with logical and snooped matrix
clocks. It is clear that snooped clocks are able to detect more
obsolete segments.}
\label{mc-vc}
\end{figure}

In Figure~\ref{racedetection}, we show the information we obtain by
the two brands of matrix clocks. It is clear that a snooped clock
allows to discard more segments than a classical logical
clock. Figure~\ref{mc-vc} compares the behaviour of logical and snooped
matrix clocks.

\subsection{Identification Phase}%%%%%%%%%%%%%%%%%%%%%%%%%%%%%%%%%%%%%%%%%%%%%%%%%%%%%%
Once a data race is found using the scheme described above we know the
address of the location and the segments containing the offending
instructions, but not the instructions themselves.  Detecting these
instructions requires another (deterministic) replay of the
program. This operation is completely automatic however and incurs no
large overhead. The replayed execution runs at full speed up to the
segments that contain the data race and from then on the
instrumentation is switched on and the offending instruction is
searched for. We are always guaranteed to find the first data race on
a particular variable.

\section{Experimental evaluation}%%%%%%%%%%%%%%%%%%%%%%%%%%%%%%%%%%%%%%%%%%%%%%%%%%%%%%

\begin{table}
\begin{small}
\begin{center}
\caption{Basis performance of RecPlay (all times in seconds)}
\label{exp1}
\begin{tabular}{l@{\,}rrrrr}
\hline
program & \cc{normal} & \multicolumn{2}{c}{record} & \multicolumn{2}{c}{replay+detect} \\
\cline{2-6}
& runtime & runtime & slowdown  & runtime & slowdown \\
\hline
cholesky      &  8.67 &  8.88 &1.024    & 721.4 & 83.2\\
fft           &  8.76 &  8.83 &1.008    &  72.8 &  8.3\\
LU            &  6.36 &  6.40 &1.006    & 144.5 & 22.7\\
radix         &  6.03 &  6.20 &1.028    & 182.8 & 30.3\\
ocean         &  4.96 &  5.06 &1.020    & 107.7 & 21.7\\
raytrace      &  9.89 & 10.19 &1.030    & 675.9 & 68.3\\
water-Nsq.    &  9.46 &  9.71 &1.026    & 321.5 & 34.0\\
water-spat.   &  8.12 &  8.33 &1.026    & 258.8 & 31.9\\
radiosity & 21.13 & 21.50 &1.018 & \multicolumn{2}{c}{data race found}\\
\hline
\multicolumn{2}{l}{geometric average}    & & 1.021 & & 30.6 \\
\hline
\end{tabular}
\end{center}
\end{small}
\end{table}      

Tables~\ref{exp1}-\ref{exp3} give an idea of the overhead caused by
\RecPlay\ during the record and the  detection phase for programs from
the SPLASH-2 benchmark suite. The experiments were performed on a Sun
multiprocessor with 4 processors. The average overhead during the
record phase is limited to 2.1\% which is small enough to keep it
switched on all the time. The automatic race detection is however very
slow: it slows down the program execution about 30 times. Remember
that the intrusion caused by this huge overhead does not alter the
execution, as this is a replayed execution.

\begin{table}
\begin{small}
\begin{center}
\caption{Performance of the data race detector}
\label{exp2}
\begin{tabular}{l@{\,}r@{\,}r@{\,}r@{\,\,}r@{\,}r}
\hline
program & \multicolumn{3}{c}{segments with memory accesses} &
\multicolumn{2}{c}{memory accesses}\\
\cline{2-4} \cline{5-6}
& \cc{created} & \multicolumn{1}{c}{max. stored}
& \cc{compared} & \cc{number} & \cc{number/s}\\
\hline
cholesky        & 13\,983  & 1\,915 (13.7\%) & 968\,154 & 121\,316\,077 & 168\,168 \\
fft             &     181  &     37 (20.5\%) &   2\,347 & 26\,463\,490 &  363\,509\\
LU              &  1\,285  &     42 \, (3.3\%) &  18\,891 & 56\,068\,996 &  388\,020\\
radix           &     303  &     36 (11.9\%) &   4\,601 & 60\,138\,828 &  328\,987\\
ocean           & 14\,150  &     47 \, (0.3\%) & 272\,037 & 29\,559\,125 &  274\,458\\
raytrace        & 97\,598  &     62 \, (0.1\%) & 337\,743 & 48\,711\,612 & 72\,069\\
water-Nsq.      &     637  &     48 \, (7.5\%) &   7\,717 & 80\,262\,966 & 249\,652\\
water-spat.     &    639   &     45 \, (7.0\%) &   7\,962 & 80\,726\,645 & 311\,927\\
radiosity       & 438\,763 & 8\,634 \, (2.0\%) & 188\,323\,337 & 275\,291\,372 & 347\,425\\
\hline
\end{tabular}
\end{center}
\end{small}
\end{table}                    

The cause of this huge overhead is explained in Table~\ref{exp2} where
the number of segments and memory accesses is shown. Apparently, for
most of the test programs, the execution of the instrumented memory
operations is the critical speed factor. For these benchmarks, this is
about 350\,000 memory operations per second. Cholesky, raytrace and
ocean do not reach that number, caused by the fact that a lot of
segments are compared during the race detection. With regard to the
number of segments stored, we see that the matrix clock algorithm
indeed succeeds in substantially reducing the number of segments that
have to be stored, limiting the memory consumption of the race
detector.

\begin{table}\small
\begin{center}
\caption{Efficiency of the ROLT mechanism}
\label{exp3}
\begin{tabular}{l@{\,}rrrrrr}
\hline
program & \cc{number of} &
\cc{size of} &
\multicolumn{2}{c}{bandwidth} \\

\cline{4-5}
& \cc{sync. op.} & \cc{trace file (b)} & \cc{bytes/s} & \cc{bits/op.}\\

\hline
cholesky  &   13\,857 &   1\,132 &    127.5 &  0.65 \\
fft       &       177 &       65 &      7.4 &  2.94 \\
LU        &    1\,275 &      134 &     20.9 &  0.84 \\
radix     &       273 &      108 &     17.4 &  3.16 \\
ocean     &   22\,981 &   6\,458 & 1\,276.3 &  2.25 \\
raytrace  &  150\,960 &  41\,416 & 4\,064.4 &  2.19 \\
water-Nsq.&       631 &      336 &     34.6 &  4.26 \\
water-spat.&      625 &      332 &     39.9 &  4.25 \\
radiosity &  524\,667 & 24\,578  & 1\,143.2 &  0.37 \\
\hline
average & & & 748.0 & 2.30 \\
\hline
\end{tabular}
\end{center}
\end{table}               

Table~\ref{exp3} shows why the overhead of recording an execution is limited to
about 2\%. This is because ROLT effectively succeeds in creating small trace
files.  The disk bandwidth needed to store the trace is never bigger than 4
kB/s, which is very low on modern machines. On the average, we only need to
store about 2.3 bits per synchronisation operation.

%%%%%%%%%%%%%%%%%%%%%%%%%%%%%%%%%%%%%%%%%%%%%%%%%%%%%%%%%%%%%%%%%%%%%%%%%%
\section{Related Work}

Although much theoretical work has been done in the field of data race
detection~\cite{adve91-05,netzer91-04,KA95,Schonberg1} few implementations for
general systems have been proposed.  Tools proposed in the past had limited
capabilities: they were targeted at programs using one semaphore~\cite{LU},
programs using only post/wait synchronisation~\cite{Netzer} or programs with
nested fork-join parallelism~\cite{KA95, Mellorcrummey4}. The tools that come
closest to our data race detection mechanism, apart from \cite{Beranek2} for a
proprietary system, is an on-the-fly data race detection mechanism for the CVM
(Concurrent Virtual Machine) system~\cite{kel_DR}. The tool only instruments the
memory references to distributed shared data (about 1\% of all references) and
is unable to perform reference identification: it will return the variable that
was involved in a data race, but not the instructions that are responsible for
the reference.

Race Frontier~\cite{choi_min} describes a similar technique as the one proposed
in this paper (replaying up to the first data race).  Choi and Min prove that it
is possible to replay up to the first data race, and they describe how one can
replay up to the race frontier.  A problem they do not solve is how to
efficiently find the race frontier. \RecPlay\ effectively solves the problem of
finding the race frontier, but goes beyond this. It also finds the cause of the
data race.

Most of the previous work, and also our \RecPlay\ tool, is based on Lamport's
{\em happened-before} relation. This relation is a partial order on all
synchronisation events in a particular parallel {\em execution}. Therefore, by
checking the ordering of all events and monitoring all memory accesses, data
races can be detected for one {\em particular program execution}.

Another approach is taken by a more recent race detector:
Eraser~\cite{eraser-tocs}. It goes slightly beyond work based on the
happened-before relation. Eraser checks that a {\em locking discipline} is used
to access shared variables: for each variable it keeps a list of locks that were
held while accessing the variable. Each time a variable is accessed, the list
attached to the variable is intersected with the list of locks currently held
and the intersection is attached to the variable.  If this list becomes empty,
the locking discipline is violated, meaning that a data race
occurred. Unfortunately, Eraser detects many false data races: as Eraser is not
based on the happened-before relation it has no timing information
whatsoever. For instance, in theory there is no need to synchronise shared
variables before multiple threads are created. The happens-before relation deals
in a natural way with the fact that threads cannot execute code `before' they
have been created but Eraser needs special heuristics to support these kind of
unlocked accesses. The support for initialisation makes Eraser {\em dependent}
on the scheduling order and therefore requires also the checking of all possible
executions for each possible input.
                                                                              
The most important problem with Eraser is however that its practical
applicability is limited as it can only process mutex synchronisation operations
and the tool fails when other synchronisation primitives are built on top of
these lock operations.

These problems with Eraser makes us believe that methods based on the
happens-before relation (like \RecPlay) are better. Contrary to Eraser, it can
only detect data races that show up in a particular program run, but it is more
general in that it knows how to deal with all common synchronisation
operations. Furthermore, it is more reliable because it never reports false data
races.

\section{Conclusions}  %%%%%%%%%%%%%%%%%%%%%%%%%%%%%%%%%%%%%%%%%%%%%%%
In this paper, we have presented \RecPlay, a practical and effective
tool for detecting data races in parallel executions.  Therefore, we
implemented a highly efficient two-level record/replay system that
traces the synchronisation operations, and uses this trace to replay
the execution. During replay, a race detection algorithm is run to
notify the programmer when a race occurs.  As such, synchronisation
races are replayed while deta races are detected. Using snooped matrix
clocks and multilevel bitmaps, we were able to limit the memory
consumption. \RecPlay\ works on running processes, and is therefore
completely independent of any compiler or programming
language. Moreover, recompiling or relinking the application is not
required.

\section*{Acknowledgements}
Michiel Ronsse is sponsored by a GOA project (12050895) from Ghent
University. Koen De Bosschere is a research associate with the Fund
for Scientific Research -- Flanders.


\begin{thebibliography}{10}

\bibitem{adve91-05}
S.V.\ Adve, M.D.\ Hill, and R.H.B.\ Netzer.
\newblock Detecting data races on weak memory systems.
\newblock In {\em Proceedings of the 18th Annual Symposium on Computer
  Architectures}, pages 234--243, May 1991.

\bibitem{KA95}
K.~Audenaert and L.~Levrouw.
\newblock Space efficient data race detection for parallel programs with
  series-parallel task graphs.
\newblock In {\em Proceedings of the third Euromicro Workshop on Parallel and
  Distributed Processing}, pages 508--515, San Remo, January 1995. IEEE
  Computer Society Press.

\bibitem{Beranek2}
Anton Beranek.
\newblock Data race detection based on execution replay for parallel
  applications.
\newblock In {\em Proceedings of CONPAR '92}, pages 109--114, Lyon, France,
  September 1992.

\bibitem{athapascan}
G.~Cavalheiro and M.~Doreille.
\newblock Athapascan: A {C}++ library for parallel programming.
\newblock In {\em Stratagem'96}, Sophia Antipolis, France, June 1996. INRIA.

\bibitem{choi_min}
Jong-Deok Choi and Sang~Lyul Min.
\newblock Race frontier: Reproducing data races in parallel-program debugging.
\newblock In {\em Proc. of the Third ACM SIGPLAN Symposium on Principles \&
  Practice of Parallel Programming}, volume~26, pages 145--154, July 1991.

\bibitem{mr-ispan}
Koen De~Bosschere and Michiel Ronsse.
\newblock Clock snooping and its application in on-the-fly data race detection.
\newblock In {\em Proceedings of the 1997 International Symposium on Parallel
  Algorithms and Networks (I-SPAN'97)}, pages 324--330, Taipei, December 1997.
  IEEE Computer Society.

\bibitem{fidge}
C.~J. Fidge.
\newblock Logical time in distributed computing systems.
\newblock In {\em IEEE Computer}, volume~24, pages 28--33. August 1991.

\bibitem{Gait2}
Jason Gait.
\newblock A probe effect in concurrent programs.
\newblock {\em Software - Practice and Experience}, 16(3):225--233, March 1986.

\bibitem{LamClock}
Leslie Lamport.
\newblock Time, clocks, and the ordering of events in a distributed system.
\newblock {\em Communications of the ACM}, 21(7):558--565, July 1978.

\bibitem{LL94-01}
Luk~J. Levrouw, Koenraad~M. Audenaert, and Jan~M. {Van C}ampenhout.
\newblock A new trace and replay system for shared memory programs based on
  {L}amport {C}locks.
\newblock In {\em Proceedings of the Second Euromicro Workshop on Parallel and
  Distributed Processing}, pages 471--478. IEEE Computer Society Press, January
  1994.

\bibitem{LU}
H.I. Lu, P.N. Klein, and R.H.B. Netzer.
\newblock Detecting race conditions in parallel programs that use one
  semaphore.
\newblock Technical report, Brown University, 1993.

\bibitem{one_sema}
Hsueh-I Lu, Philip~N. Klein, and Robert H.~B. Netzer.
\newblock Detecting race conditions in parallel programs that use one
  semaphore.
\newblock Workshop on Algorithms and Data Structures (WADS), Montreal, August
  1993.

\bibitem{Mattern}
Friedemann Mattern.
\newblock Virtual time and global states of distributed systems.
\newblock In Cosnard, Quinton, Raynal, and Roberts, editors, {\em Proceedings
  of the Intl. Workshop on Parallel and Distributed Algorithms}, pages
  215--226. Elsevier Science Publishers B.V., North-Holland, 1989.

\bibitem{Mellorcrummey4}
John~M. Mellor-Crummey.
\newblock On-the-fly detection of data races for programs with nested fork-join
  parallelism.
\newblock In {\em Proceedings of Supercomputing '91}, pages 24--33, November
  1991.

\bibitem{netzer91-04}
R.H.B.\ Netzer and B.P.\ Miller.
\newblock Improving the accuracy of data race detection.
\newblock In {\em Proceedings of the 1991 Conference on the Principles and
  Practice of Parallel Programming}, April 1991.

\bibitem{Netzer}
Robert H.~B.\ Netzer and Barton~P.\ Miller.
\newblock On the complexity of event ordering for shared-memory parallel
  program executions.
\newblock International Conference on Parallel Processing, pages 93--97, August
  1990.

\bibitem{Netzer93}
Robert~H.B. Netzer.
\newblock Optimal tracing and replay for debugging shared-memory parallel
  programs.
\newblock In {\em Proceedings ACM/ONR Workshop on Parallel and Distributed
  Debugging}, pages 1--11, May 1993.

\bibitem{Netzer_DR}
Robert~H.B.\ Netzer and Barton~P.\ Miller.
\newblock What are race conditions? some issues and formalizations.
\newblock {\em ACM Letters on Programming Languages and Systems}, March 1992.

\bibitem{kel_DR}
Dejan Perkovic and Peter~J. Keleher.
\newblock {Online Data-Race Detection via Coherency Guarantees}.
\newblock pages 47--57, Seattle, October 1996.
\newblock The Second Symposium on Operating Systems Design and Implementation
  (OSDI '96) Proceedings.

\bibitem{raynal96}
Michel Raynal and Mukesh Singhal.
\newblock Logical clocks: Capturing causality in distributed systems.
\newblock {\em IEEE Computer}, pages 49--56, February 1996.

\bibitem{MR-08}
M.~Ronsse and L.~Levrouw.
\newblock On the implementation of a replay mechanism.
\newblock In L.~Bouge, P.~Fraigniaud, A.~Mignotte, and Y.~Robert, editors, {\em
  Proceedings of EuroPar `96}, volume 1123 of {\em LNCS}, pages 70--73.
  Springer-Verlag, Lyon, August 1996.

\bibitem{MR95-03}
M.~Ronsse, L.~Levrouw, and K.~Bastiaens.
\newblock Efficient coding of execution-traces of parallel programs.
\newblock In J.~P.\ Veen, editor, {\em Proceedings of the ProRISC / IEEE
  Benelux Workshop on Circuits, Systems and Signal Processing}, pages 251--258.
  STW, Utrecht, March 1995.

\bibitem{MR_JITI}
Michiel Ronsse and Koen De~Bosschere.
\newblock {JiTI: Tracing Memory References for Data Race Detection}.
\newblock In E.~D'Hollander, F.J. Joubert, and U.~Trottenberg, editors, {\em
  Proceedings of ParCo97: Parallel Computing: Fundamentals, Applications and
  New Directions}, volume~12 of {\em Advances in Parallel Computing}, pages
  327--334, Bonn, February 1998. North Holland.

\bibitem{mr-tocs}
Michiel Ronsse and Koen De~Bosschere.
\newblock Recplay: A fully integrated practical record/replay system.
\newblock {\em ACM Transactions on Computer Systems}, 17(2):133--152, May 1999.

\bibitem{sarin}
S.K. Sarin and L.~Lynch.
\newblock Discarding obsolete information in a replicated data base system.
\newblock In {\em IEEE Transactions on Software Engineering}, volume~SE, pages
  39--46. January 1987.

\bibitem{eraser-tocs}
Stefan Savage, Michael Burrows, Greg Nelson, Patrick Sobalvarro, and Thomas
  Anderson.
\newblock Eraser: A dynamic data race detector for multithreaded programs.
\newblock {\em ACM Transactions on Computer Systems}, 15(4):391--411, November
  1997.

\bibitem{Schonberg1}
Edith Schonberg.
\newblock On-the-fly detection of access anomalies.
\newblock {\em Proceedings of the SIGPLAN '89 Conference on Programming
  Language Design and Implementation, published in ACM SIGPLAN Notices},
  24(7):285--297, July 1989.

\bibitem{locklint}
SunSoft.
\newblock {lock\_lint User's Guide}, 1994.

\bibitem{splash2}
Steven~Cameron Woo, Moriyosho Ohara, Evan Torrie, Jaswinder~Pal Singh, and
  Anoop Gupta.
\newblock The {SPLASH-2} programs: Characterization and methodological
  considerations.
\newblock In {\em Proc. of the 22nd Annual International Symposium on Computer
  Architecture}, pages 24--36, June 1995.

\bibitem{wuu}
G.T.J. Wuu and A.J. Bernstein.
\newblock Efficient solutions to the replicated log and dictionary problems.
\newblock Proc. 3rd ACM Symp. Principles Distributed Computing, pages 233--242,
  New York, 1984. ACM Press.

\end{thebibliography}
\end{document}